\documentstyle[prl,aps,epsf]{revtex}
\draft
\begin{document}
\twocolumn[\hsize\textwidth\columnwidth\hsize\csname @twocolumnfalse\endcsname

\title{Critical Depinning Force and Vortex Lattice Order in
Disordered Superconductors} 
\author{C. J. Olson and C. Reichhardt}
\address{Department of Physics, University of California, Davis, California
95616.}

\author{S. Bhattacharya} 
\address{NEC Research Institute, 4 Independence Way, Princeton,
New Jersey 08540.} 

\date{\today}
\maketitle
\begin{abstract}
We simulate the 
ordering of vortices and its effects on the critical current in
superconductors with varied vortex-vortex interaction strength and varied
pinning strengths for a two-dimensional system.
For strong pinning the vortex lattice is always disordered
and the critical depinning force only weakly increases with decreasing
vortex-vortex interactions. For weak pinning the vortex lattice is 
defect free until the vortex-vortex interactions have been
reduced to a low value, when defects 
begin to appear with a simultaneous
rapid increase in the critical depinning force.
In each case the depinning force shows a 
maximum for non-interacting vortices.
The
relative height of the peak increases and the peak width
decreases for decreasing pinning strength
in excellent agreement with experimental trends associated with the peak
effect.
We show that
scaling relations exist 
between the distance between defects in the vortex lattice and the
critical depinning force.
\end{abstract}

\vskip2pc]
\narrowtext

\section{Introduction}
A large class of condensed matter systems can be represented as an elastic
lattice interacting with a quenched random substrate.  
Vortices in type II superconductors,
which can be pinned by random defect sites in the material, are a 
particularly ideal example of such a system since the substrate and lattice 
interactions can be readily tuned experimentally.
In this case, a rich variety of static 
\cite{Blatter1,Higgins2,Sahni3,Ghosh4,Ling5,Banerjee6} 
and dynamical phases \cite{Higgins2,Driven7,Henderson8} can 
occur due to the competition between the 
repulsive vortex-vortex interactions, which tend to order the
vortex lattice, and the attractive vortex-pin interactions,
which tend to disorder the vortex lattice. 
Unlike the early studies of collective pinning of a defect-free elastic
lattice, recent experimental and theoretical work point to the importance
of topological defects in pinned vortex matter.
If the pinning in the sample is sufficiently strong,
the vortex lattice will be highly defected; 
conversely, in a sample with weak pinning
the vortex lattice can be relatively ordered and contain only a few defects.  
In contrast with the purely elastic model, however, the relationship
between the critical force and the density of flux line defects
is not known.
Moreover, applications of superconducting materials require that the
critical current be high, and so
a key question is how the number of defects in the vortex lattice can 
affect the critical current.         

In a given sample, 
i.e., with a given realization of quenched disorder,
the relative strength of the pinning can be increased
by lowering the vortex-vortex interaction strength, 
i.e., by softening the lattice.
Vortex-vortex interactions become weak at low applied fields, when vortices
are dilute,
as well as very 
near $H_{c2}$ and $T_{c}$, 
where the induction of the vortices overlap significantly.
In the latter regime one encounters the peak effect: a rapid increase
in the critical current before it collapses to zero at the
superconducting - normal phase boundary.
At these two extreme field values, the pinning interaction dominates
and the vortex lattice is expected to be highly disordered.
Indeed, direct evidence for vortex lattice disordering 
through the peak 
effect regime has been observed in neutron scattering 
measurements \cite{gammel9}. 
A reentrant disordering has been found in
the low field regime where the vortex lattice
again softens \cite{Ghosh4}. 
Bitter decoration experiments have also provided evidence for the
disordering of the vortex lattice at low fields \cite{Grier9a}.  
In each of these cases the critical current $J_c$ is high due to the
dominance of pinning.

Experiments on 
superconducting samples with different amounts of pinning have revealed 
some systematic trends.
The peak effect is much sharper and more pronounced
in cleaner samples \cite{Banerjee6,Ling5,deGroot10}, 
while in samples with stronger pinning, 
the width of the peak region increases and 
{\it the relative valley-to-peak height decreases}, 
although the overall critical current is higher \cite{Banerjee6,deGroot10}.
In samples with the strongest pinning,
$J_{c}$ is high through most of the 
applied field range, and the peak is completely absent.
A possible interpretation of this behavior connects vortex lattice
defects with high critical currents.
A detailed understanding of the observed effects as a possible
transition/crossover between regimes with low and high critical
currents, i.e., between, say, a Bragg glass-like ordered phase with
few defects and a Vortex glass/ pinned liquid-like disordered vortex
phase is still lacking.  A key reason for this uncertainty is the
absence of a systematic connection between the number of defects in
the vortex lattice, the softness of the lattice, and the critical current.

In this work, we
directly examine the effect of vortex lattice softness on the critical current 
through a series of numerical simulations of samples with different
pinning strengths.
For strong pinning the vortex lattice is highly defective for the entire
range of vortex-vortex interactions investigated. The critical depinning
force, which is proportional to the critical current,  
only weakly increases with decreasing vortex-vortex
interaction strength, reaching a maximum 
for non-interacting vortices.
For the weakest pinning strengths, on the other hand, the
lattice is almost defect-free over a large range of vortex-vortex
interaction strengths.
When defects begin to appear in the softest vortex lattices,
the critical depinning force increases sharply 
from its low value to a peak at zero vortex-vortex interaction. 
The behavior for intermediate pinning is intermediate between these
extremes.
We find that the relative height of the critical current peak
increases for decreasing pinning while the width of the peak
decreases.  Thus, the most pronounced peaks occur for the weakest 
pinning. 
These results are in excellent agreement with the experimental trends
described above.
We discuss how these results can be 
connected to the behavior of the peak effect in superconductors with
varying pinning strength.       

\section{Simulation}
We consider a 2D slice of a system of superconducting vortices interacting
with a random pinning background.  The applied magnetic field 
${\bf H}=H{\bf \hat{z}}$ is perpendicular to our sample, and we use
periodic boundary conditions in $x$ and $y$.  The $T=0$ 
overdamped equation of motion 
for a vortex $i$ is 
\begin{equation}
{\bf f}_{i} = \eta \frac{d {\bf r}_{i}}{dt} = 
{\bf f}_{i}^{vv} + {\bf f}_{i}^{vp} + {\bf f}_{d} = \eta {\bf v}_{i}
\ ,
\end{equation}
where ${\bf v}_{i}$ is the velocity of vortex $i$ and 
$\eta = 1$ is the damping coefficient.
The total force on vortex
$i$ from the other vortices is 
$f_{i}^{vv} = \sum_{j = 1}^{N_{v}}A_{v} \ f_{0}
K_{1}(|{\bf r}_{i} - {\bf r}_{j}|/\lambda){\bf \hat{r_{ij}}}$,
where ${\bf r}_{i}$ is the position of vortex $i$, $\lambda$ is
the penetration depth, $f_0 = \Phi_0^{2}/8\pi^2 \lambda^3$,
the prefactor $A_v$ is used to vary the vortex lattice softness, and
$K_{1}$ is the modified Bessel function.  
The pinning ${\bf f}_{i}^{vp}$ 
is modeled as randomly placed attractive parabolic wells of
radius $r_{p}=0.15\lambda$ with 
$f_{i}^{vp} = (f_{p}/r_{p})(|{\bf r}_{i}-{\bf r}_{k}^{(p)}|)
\ \Theta(r_p - |{\bf r}_{i} - {\bf r}_{k}^{(p)}|) 
{\hat {\bf r}}_{ik}^{(p)}$, 
where ${\bf r}_{k}^{(p)}$ is the location of pin
$k$, $f_{p}$ is the maximum pinning force, 
which is varied from $0.1 f_0$ to $3.0 f_0$,
$\Theta$ is the
Heavside step function and ${\hat {\bf r}}_{ik}^{(p)} = 
({\bf r}_{i} - {\bf r}_{k}^{(p)})/| {\bf r}_i - {\bf r}_{k}^{(p)}|$.
The pin density is $n_p = 3.0/\lambda^{2}$ and the vortex density
is $n_v = 0.75/\lambda^{2}$.  We simulate a $36\lambda \times 36\lambda$
system containing $N_v = 864$ vortices and $N_p = 3887$ pins.
We initialize the vortex positions by performing
simulated annealing, starting from a high
temperature and slowly cooling to $T=0$. This method of preparing the
lattice is similar to field cooled experiments.
To identify the depinning force $f_c$ we apply a slowly
increasing uniform driving force ${\bf f}_{d}$ on the vortices in the 
$x$ direction,  which
would correspond to a 
Lorentz force from an applied current ${\bf J}=J{\bf \hat{y}}$.
For each drive increment we measure the average vortex 

\begin{figure}
\center{
\epsfxsize = 3.5 in
\epsfbox{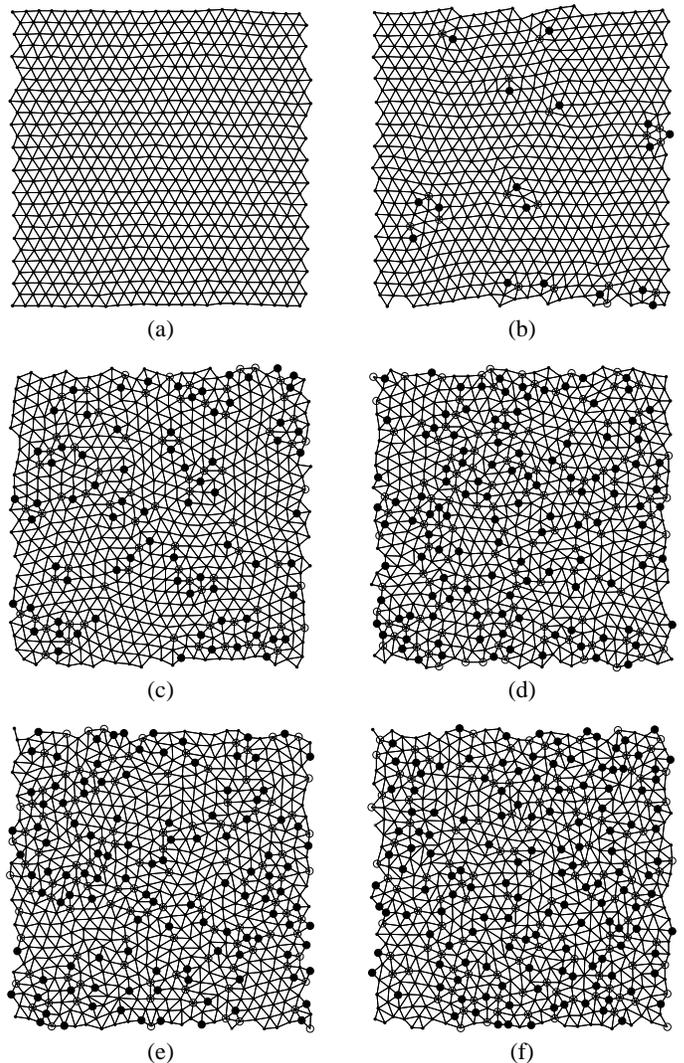}}
\caption{
The Delaunay triangulation of the vortex lattice for decreasing 
vortex-vortex interaction strength $A_{v}$ in a sample
with $f_{p} = 0.25f_0$. 
Dark circles indicate 5-fold coordinated vortices; open circles
indicate 7-fold coordinated vortices.
(a) $A_{v} = 4.0$, 
(b) $3.0$, (c) $2.0$, (d) $1.0$, (e) $0.75$ and (f) $0.50$.  
The vortex lattice is relatively ordered in (a).
In (b) more defects appear.  (c) and (d) show 
different domains of vortex orientation.
In (e) and (f) the vortex lattice is highly disordered.}
\label{fig:delaunay}
\end{figure} 

\hspace{-13pt}
velocity in 
the direction of drive, 
$V_{x} = (1/N_v) \sum_{1}^{N_{v}}v_x$.  
The 
$f_{d}$ versus $V_{x}$ curve corresponds experimentally to
a $V(I)$ curve. The depinning force $f_c$ is defined as the drive
at which $V_{x} > 0.03$.

\section{Vortex Order and Pinning for Varied Vortex-Vortex Interaction 
Strength}

We first consider the effect of the vortex lattice softness on the 
stationary vortex lattice.  
In Fig.~\ref{fig:delaunay} 
we show the Delaunay triangulation for a system with $f_{p} = 0.25f_0$ 
for decreasing vortex-vortex interaction $A_{v}$, after the 

\begin{figure}
\center{
\epsfxsize = 3.5 in
\epsfbox{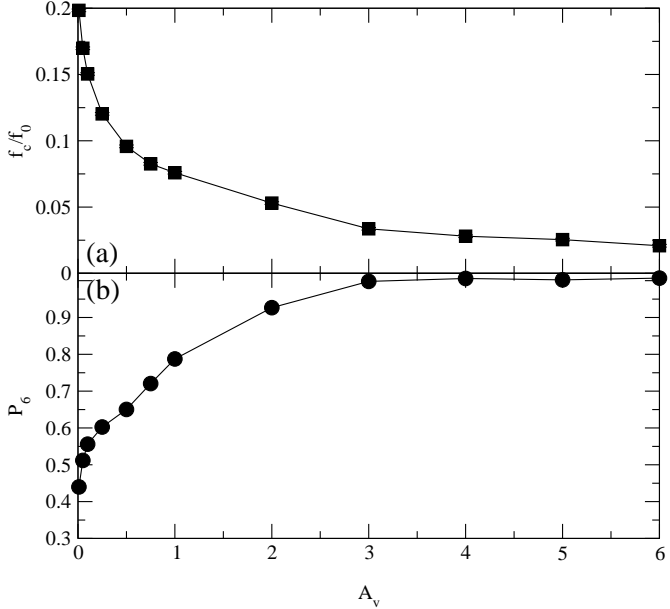}}
\caption{
(a) The critical depinning force $f_{c}$ versus $A_{v}$ 
for a sample with $f_p = 0.25 f_0$.
(b) The fraction of six-fold coordinated vortices $P_6$ versus $A_{v}$. As the
vortex-vortex interaction is lowered $f_{c}$ increases while $P_6$
decreases.}
\label{fig:jcp6}
\end{figure} 

\hspace{-13pt}
lattice has
been annealed and with no driving force applied. 
Defect sites in the vortex lattice are indicated by circles.
In Fig.~\ref{fig:delaunay}(a) for 
$A_{v} = 4.0$, the vortex lattice contains no defects.
In Fig.~\ref{fig:delaunay}(b) for $A_{v} = 3.0$
a small number of 5-7 defect pairs appear. 
In Fig.~\ref{fig:delaunay}(c), at $A_{v}= 2.0$ the vortex lattice is
considerably disordered but some domains of order are still
present.  For softer vortex lattices in 
Fig.~\ref{fig:delaunay}(d-f), the vortex lattice becomes progressively
more disordered by the underlying random pinning.
Since the simulation is for a two-dimensional system, we expect the
system to be defective for any pinning strength.  The defect-free
situation likely results from the finite size of the sample.

By applying a transport current to the annealed lattices, we determine
the critical depinning force $f_{c}$ that must be applied before the
vortices begin to move.
In Fig.~\ref{fig:jcp6}(a) we plot $f_{c}$ versus $A_{v}$ for a sample
with $f_{p}=0.25f_0$.  For comparison, we measure the amount of order
in the lattice before depinning using the Delaunay triangulation, and
in Fig.~\ref{fig:jcp6}(b) we plot $P_{6}$, the fraction of
six-fold coordinated vortices, versus $A_{v}$. Here 
$P_{6} = 1.0$ indicates a perfect
triangular lattice. As $A_{v}$ is lowered $f_{c}$ increases 
and simultaneously the order in the lattice, $P_{6}$, decreases. 
The maximum value of $f_{c}$ occurs
at $A_{v} \sim 0$ which coincides with the minimum value of $P_{6}$. Thus the 
softer lattices with low values of $A_v$ and large amounts of
disorder are more strongly pinned than stiffer, more ordered lattices.

\begin{figure}
\center{
\epsfxsize = 3.5 in
\epsfbox{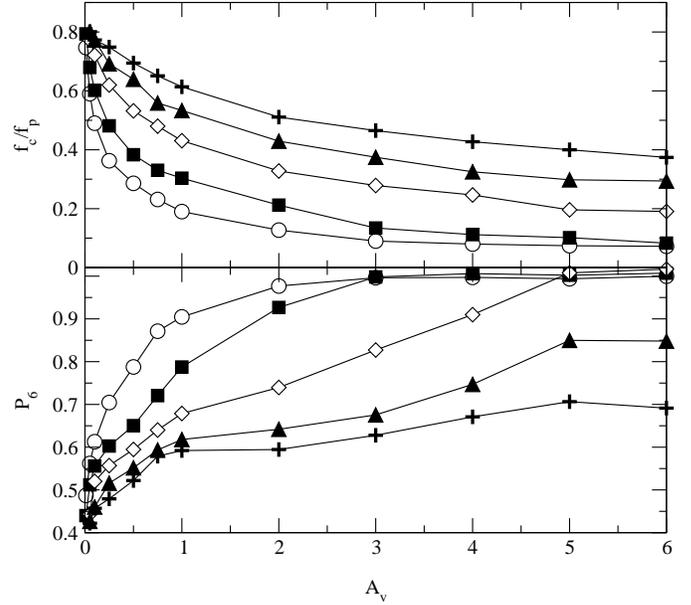}}
\caption{
Scaled critical depinning force and $P_6$ versus $A_v$ for samples
with different pinning strengths.  Open circles: $f_p = 0.10 f_0$;
filled squares: $f_p = 0.25 f_0$; open diamonds: $f_p = 0.75 f_0$;
filled triangles: $f_p = 1.5 f_0$; plus signs: $f_p = 3.0 f_0$.
(a) The plot of $f_{c}/f_{p}$ versus $A_{v}$ shows that the peak at low 
$A_{v}$ becomes sharper for weakening pinning strength $f_{p}$. 
(b) The corresponding $P_{6}$ values show that for large $f_{p} = 3.0$ 
the vortex lattice is disordered over the entire range of $A_{v}$.}
\label{fig:varyfp}
\end{figure} 

\section{Height and Width of Critical Depinning Force Peak}

The shape and magnitude of the critical current peak at $A_v = 0$ are
affected by the strength of the pinning in the sample.  To demonstrate
this, in Fig.~\ref{fig:varyfp}(a) we plot $f_{c}/f_{p}$ versus $A_{v}$.
Scaling the curves with $f_p$ in this way causes all of the curves to
approach $f_c/f_p = 1$ at $A_v = 0$.
We find that the relative height of the peak increases as the pinning
becomes weaker.  For example,
comparing $f_p = 0.25$ (bottom curve) with $f_p = 3.0$ (top curve),
the critical current $f_c$ increases by a factor of 8
from $f_c = 0.1$ at $A_v = 6.0$ to 
$f_c = 0.8$ at $A_v = 0.01$ in the weakly pinned sample, 
whereas in the strongly pinned sample, $f_c$ increases only by a
factor of 2 from $f_c = 0.38$ at $A_v = 6.0$ to $f_c = 0.8$ at 
$A_v = 0.01$.  The peak not only becomes higher for weaker pinning,
but also it becomes much narrower as can be seen by comparing the 
widths of the peaks in Fig.~\ref{fig:varyfp}(a).
As shown in Fig.~\ref{fig:varyfp}(b), in the strongly pinned
samples such as $f_p = 3.0f_0$ 
with broad low peaks, $P_6$ is low over the entire range
of $A_v$, indicating that the vortex lattice is always strongly
disordered by the pinning.  In contrast, for weakly pinned samples
such as $f_p = 0.25f_0$, $P_6$ drops a large amount from $P_6\approx 1$
at large $A_v$, indicating a nearly perfect 

\begin{figure}
\center{
\epsfxsize = 3.5 in
\epsfbox{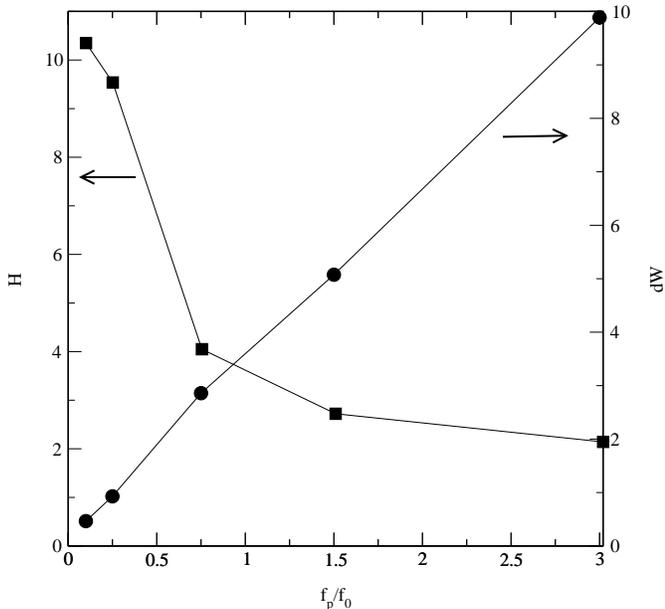}}
\caption{Squares:
the relative height $H = f_{c}(A_v=0.01)/f_{c}(A_v=6.0)$
of the peak in $f_c$ for samples with different pinning strength
$f_p$.  Circles: The width of the peak $dW$ at half-maximum
as a function of $f_p$.  The 
peak becomes sharper and more pronounced for weaker pinning.}
\label{fig:width} 
\end{figure}

\hspace{-13pt}
lattice, to $P_6\approx 0.4$
at the lowest $A_v$, indicating a large amount of disorder in the lattice.

In Fig.~\ref{fig:width} we show
explicitly how the relative peak height, 
$H = f_{c}(A_{v} = 0.01)/f_{c}(A_v = 6.0)$, and the width at 
half maximum, $dW$, vary with pinning strength $f_{p}$. 
Here we see that the relative
strength and sharpness of the peak increase with weaker pinning. The
width $dW$ depends linearly on $f_{p}$, 
while the relative height $H$ shows a nonlinear
increase that can not be fitted by a simple functional form. 

The results in Fig.~\ref{fig:varyfp} and Fig.~\ref{fig:width}
suggest that in clean superconductors, 
where the vortex lattice softness has a much more significant impact
on the effectiveness of the pinning,
the critical current enhancements should be
both sharper and of larger relative height than in
strongly pinned superconductors.  This behavior agrees with the results
of recent experiments \cite{Ling5,Banerjee6} on the peak effect. 
Our results also suggest that the peak can be made arbitrarily 
sharp simply by adjusting  the disorder strength, and that 
such large peaks in the critical current 
must be accompanied by the proliferation of defects. 

\section{Critical Depinning Force Dependence on Defect Density} 
\begin{figure}
\center{
\epsfxsize = 3.5 in
\epsfbox{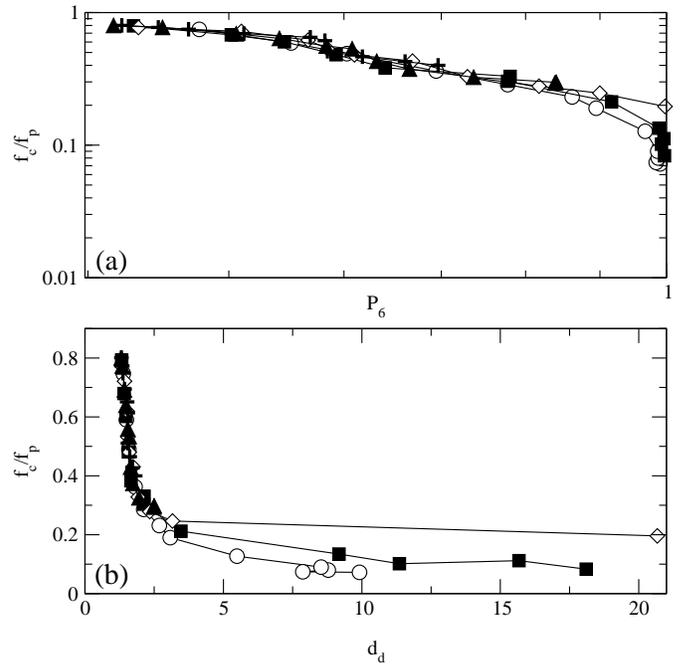}}
\caption{
(a) Scaled critical depinning force $f_{c}/f_{p}$ 
versus $P_{6}$. 
(b) $f_{c}/f_{p}$ versus the distance 
between defects $d_{d}$ showing a 
collapse for $d_{d} < 3.0$ coinciding with a rapid increase in $f_{c}$.  }
\label{fig:scale}
\end{figure}

Using the data obtained by varying $A_v$, we can determine
the relationship between the critical current and the density
of defects.
In Fig.~\ref{fig:scale}(a) we plot $f_{c}/f_{p}$ 
versus $P_{6}$, showing that the curves for
samples with different 

\hspace{-13pt}
pinning strengths roughly
collapse. 
At lower values of $P_{6}$ when the vortex lattice becomes
defected, 
the critical
current increases rapidly with the onset of defects.  
In Fig.~\ref{fig:scale}(b) 
we plot $f_{c}/f_{p}$ versus the distance between 
vortex lattice defects, 
$d_{d} = 1/\sqrt{1.0 - P_{6}}$.
In samples with weak pinning
there is a range of high $A_v$ values that produce well ordered
vortex lattices.  In this regime, when there are very few defects in
the vortex lattice which are far apart, we find that
the $f_{c}/f_{p}$ values do not collapse, and that the rate at
which $f_{c}/f_{p}$ changes with $d_d$ is slow.
For $d_{d} < 3.0$, or when the lattice contains approximately 10\% 
or more defects,
all the curves collapse and 
$f_{c}/f_{p}$ increases rapidly with decreasing $d_d$. 
The behavior of $f_{c}$ versus $d_d$
suggests that there is a critical distance between defects below which 
$J_{c}$ begins to rapidly increase. 

\begin{figure}
\center{
\epsfxsize = 3.5 in
\epsfbox{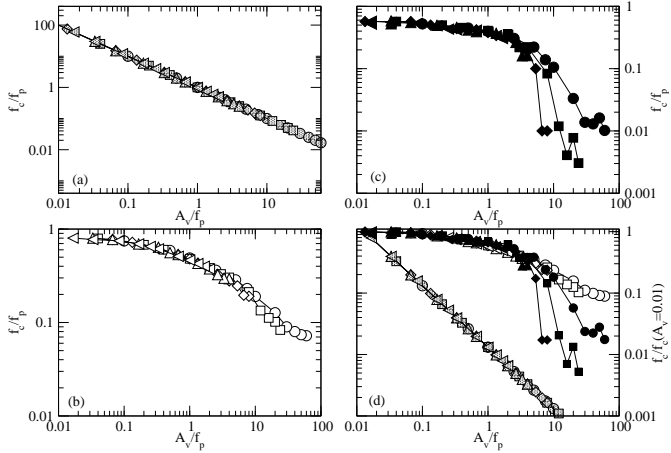}}
\caption{(a) Shaded symbols: $f_c/f_{p}$ as predicted by LO theory.
(b) Filled symbols: $f_c/f_{p}$ computed 
from the distance between dislocations.
(c) Open symbols: $f_c/f_{p}$ measured in the simulations.
(d) All three methods of obtaining $f_c$ are scaled by the value at
$A_v = 0.01$, and plotted together.
For all panels, circles: $f_p = 0.10 f_0$;
squares: $f_p = 0.25 f_0$; diamonds: $f_p = 0.75 f_0$;
triangle up: $f_p = 1.5 f_0$; triangle left: $f_p = 3.0 f_0$.
}
\label{fig:scale2}
\end{figure}

We can compare the dependence of the critical depinning force on the
defect density with the collective pinning results of the
Larkin-Ovchinnikov (LO) theory \cite{Larkin11}.  
In Fig.~\ref{fig:scale2}(a) we plot
$f_c$ as predicted by LO, using $f_c \propto f_p^2 / A_v$.  
Fig.~\ref{fig:scale2}(b) shows
$f_c$ computed from the distance between dislocations,
using $f_c \propto f_{p}/d_d$.
The value of $f_c$ measured in the simulations
is shown in Fig.~\ref{fig:scale2}(c).
We compare $f_c$ obtained by the LO and dislocation distance methods
with the actual $f_c$ in Fig.~\ref{fig:scale2}(c).
The agreement with the LO prediction is poor, but that with the
dislocation distance is good out to values $A_v/f_p \sim 5$,
beyond which $d_d$ becomes of order the system size.

\section{Conclusion}
We have investigated the
dependence of the critical depinning force $f_{c}$ and 
vortex lattice topological order $P_6$ on the vortex lattice 
rigidity for different values of pinning strength.  The study is
restricted to the two dimensional case which is appropriate for vortex
lattices where the longitudinal correlation length is large enough to
be effectively given by the sample thickness.
In all cases,
$f_{c}$ increases with decreasing vortex-vortex interaction strength,
reaching a maximum for non-interacting vortices, as expected.
Although for very weak pinning and strong vortex-vortex interaction, 
the lattice is relatively defect-free, 
for strong pinning the lattice is defective for the entire vortex
interaction range investigated.
These results are in accord with previous studies.
We have, in addition, studied the systematic variation of $f_c$
with varying pinning and inter-vortex interaction.  The results
show that the pinning force variation departs significantly from the
expectations of an elastic picture.  The increase in $f_c$ towards the
maximum occurs very rapidly with decreasing interaction or increasing
pinning. The relative height increases while the width decreases 
with decreasing pinning strength, strongly remniscent of the peak
effect phenomenon.  Furthermore, a pseudo-Larkin picture where a plastic
length, the distance between topological defects, replaces the elastic
correlation length, provides a good account of the variation of $f_c$.
We also find evidence that there is a critical value of the plastic
length, typically spanning 3-5 vortices on each side, at which the
rapid crossover occurs.
These results    
should be relevant to experiments in very weak pinning (quasi-2D) flux
lattices, as well as to the general systems of two-dimensional lattices
with quenched disorder where the lattice interaction can be tuned.

\end{document}